\documentclass[fleqn,usenatbib]{mnras}

\usepackage{newtxtext,newtxmath}

\usepackage[T1]{fontenc}
\usepackage{ae,aecompl}

\DeclareRobustCommand{\VAN}[3]{#2}
\let\VANthebibliography\thebibliography
\def\thebibliography{\DeclareRobustCommand{\VAN}[3]{##3}\VANthebibliography}

\usepackage{graphicx}	
\usepackage{amsmath}	
\usepackage{xcolor}
\colorlet{RED}{red}
\usepackage{soul}
\usepackage{physics}
\usepackage{bm}

\usepackage{hyperref}
\hypersetup{linkcolor=cyan,citecolor=blue,filecolor=cyan,urlcolor=blue}
\usepackage{color}
\title[Ballistic Surfing Acceleration in Galaxy Clusters]{Ballistic Surfing Acceleration as a Coherent Mechanism for Electron Acceleration in Galaxy Cluster Shocks}

\author[J.-H. Ha \& K. Stasiewicz]{
Ji-Hoon Ha,$^{1}$\thanks{E-mail: jhha@kasi.re.kr}
Krzysztof Stasiewicz,$^{2}$
\\
$^{1}$Korea Astronomy and Space Science Institute (KASI), 34055 Daejeon, Republic of Korea\\
$^{2}$Space Research Centre, Polish Academy of Sciences, Bartycka 18A, PL-00-716 Warsaw, Poland
}

\date{Accepted 2026 XXX. Received YYY; in original form ZZZ}
\pubyear{2026}

\begin{document}
\label{firstpage}
\pagerange{\pageref{firstpage}--\pageref{lastpage}}
\maketitle

\begin{abstract}
Radio relics in merging galaxy clusters are widely interpreted as synchrotron emission from relativistic electrons accelerated at large-scale shocks. However, the efficiency of diffusive shock acceleration (DSA) is expected to be suppressed in the low-Mach-number, weakly turbulent environments of cluster mergers; furthermore, recent theoretical insights suggest that DSA may not constitute a viable physical mechanism for such environments. In this work, we investigate ballistic surfing acceleration (BSA) as an electrodynamically grounded alternative for electron energization that bypasses the need for prescribed diffusion coefficients. We formulate BSA under typical cluster shock conditions, deriving the balance between coherent acceleration by the convective electric field and radiative losses from synchrotron and inverse-Compton cooling. This equilibrium defines the maximum reachable electron energy and constrains the resulting steady-state spectrum. By forward-modeling the associated synchrotron emission and comparing it with integrated radio observations of the `Sausage' (CIZA J2242.8+5301) and `Toothbrush' (1RXS J0603.3+4214) relics, we find that the observed spectral curvature and high-frequency steepening are consistent with BSA-limited energies, provided that active acceleration involves only a minute participation fraction ($10^{-9}$--$10^{-8}$) of the radiating electrons. Despite this high selectivity, BSA effectively produces Lorentz factors of $\gamma \sim 10^4$--$10^5$. Our results suggest that radio relics serve as prime astrophysical laboratories for probing coherent acceleration, with the BSA framework providing a robust and physically consistent explanation for electron energization in cluster shocks.
\end{abstract}


\begin{keywords}
shock waves  -- acceleration of particles -- radiation mechanisms: Non-thermal -- galaxies: clusters: intracluster medium
\end{keywords}



\section{Introduction}
Radio relics are large-scale, diffuse synchrotron sources observed predominantly in the peripheral regions of merging galaxy clusters \citep[e.g.,][and references therein]{vanWeeren2010,Bruggen2012, Stroe2016,vanWeeren2017,vanWeeren2019, Rajpurohit2021}. They are typically elongated, highly polarized, and characterized by steep radio spectra, strongly suggesting an origin associated with merger-driven shock waves propagating through the intracluster medium (ICM). The relativistic electrons responsible for the observed synchrotron emission are generally inferred to have Lorentz factors of order $\gamma \sim 10^4$--$10^5$ \citep[e.g.,][]{Hoeft2007, Siemieniec-Ozieblo2011, Pinzke2013}.

The standard theoretical framework employed to interpret radio relics is diffusive shock acceleration (DSA) \citep[e.g.,][]{Bell1978, Blandford1978, Drury1983}, in which particles gain energy through repeated shock crossings mediated by scattering off magnetic turbulence. Within this paradigm, the observed radio spectral index is commonly used to infer the shock Mach number.
However, the application of DSA to galaxy cluster merger shocks remains an active area of investigation. Cluster shocks are typically characterized by low Mach numbers ($\mathcal{M} \lesssim 3$), weak magnetic fields, and uncertain levels of turbulence—conditions under which efficient electron injection and acceleration may be more difficult to achieve. Semi-analytic and numerical studies suggest that the acceleration efficiency decreases toward lower Mach numbers in this regime \citep[e.g.,][]{Kang2013,Ryu2019, Kang2019, Ha2021, Ha2022}, although the extent of this reduction continues to be explored.
In addition, DSA models often rely on parameterized descriptions of particle diffusion and pitch-angle scattering whose microphysical origins are still being investigated, particularly in the tenuous intracluster medium (ICM).

Several physical mechanisms have been proposed to improve electron acceleration at weak cluster shocks within the broader diffusive framework. One possibility is that pre-existing turbulence in the upstream medium modifies particle transport and scattering near the shock, thereby modulating the efficiency of electron acceleration \citep[e.g.,][]{Trotta2021,Guo2021,Ha2025}. Recent studies suggest that externally driven or inherited turbulence can, in principle, lower the effective injection threshold and modify the spectral properties of accelerated particles, although the required turbulence levels and coherence scales remain uncertain in the intracluster medium \citep[e.g.,][]{Fujita2015, Dominguez2024}.
Another widely discussed scenario involves the re-acceleration of pre-existing suprathermal or fossil electron populations, which can partially bypass the injection barrier and produce observable radio emission even at low Mach numbers \citep[e.g.,][]{Kang2011,Kang2012,Pinzke2013,vanWeeren2017}. These scenarios highlight the diversity of pathways through which relativistic electrons may be generated in cluster environments.
At the same time, such mechanisms depend on environmental factors—including the level of upstream turbulence and the availability of nonthermal seed particles—that are expected to vary across clusters and merger histories. For example, radio relics are predominantly located in the outskirts of merging systems, where the number density of active galactic nuclei may be lower than in cluster cores \citep[e.g.,][]{Feretti2012}, potentially influencing the supply of fossil electrons.
Taken together, these considerations point toward a complex acceleration landscape in which multiple processes may contribute, motivating the exploration of alternative channels that could replace diffusive acceleration mechanisms in galaxy cluster shocks.

Diffusion, whether occurring in velocity space and/or in configuration space, is a relaxation process that drives the distribution function toward equilibrium and does not necessarily lead to systematic particle energization. In contrast, ballistic surfing acceleration (BSA) \citep[][]{Stasiewicz2025}, which arises naturally from the presence of magnetic-field gradients in collisionless shocks, provides a direct acceleration mechanism for particles whose gyroradii exceed the width of the magnetic ramp in a shock.
In this framework, particle energization results from work done by the convection electric field generated by plasma flow across the magnetic field, as observed in the shock reference frame. Particles gain more energy in the upstream region than they lose in the downstream region, leading to systematic acceleration when averaged over a gyroperiod. Processes occurring within and around the shock — such as resonant wave–particle interactions or turbulent diffusion — do not contribute significantly to the acceleration of high-energy particles gyrating across the shock, but instead act primarily to isotropize their distribution.

A key feature of BSA is its strong dependence on shock geometry: the acceleration efficiency is expected to be highest at perpendicular shocks, where the motional electric field reaches its maximum amplitude, and to decrease toward the parallel limit. Importantly, cosmological simulations of large-scale structure formation suggest that quasi-perpendicular shocks are common, accounting for roughly $\sim$70\% of merger-driven and cosmological shocks  \citep[e.g.,][]{Wittor2017,Banfi2020,Ha2020}.
This trend has been reported both in large cosmological volumes and in cluster-focused simulations designed to reproduce radio relic environments, which similarly find that relic-hosting shocks are frequently quasi-perpendicular \citep[e.g.,][]{Roh2019}. Consequently, the geometric conditions favorable for BSA appear to be widespread in galaxy cluster merger shocks.

In this paper, we investigate whether BSA can operate efficiently in galaxy cluster merger shocks and contribute to the relativistic electron populations responsible for radio relic emission. Focusing on typical cluster shock conditions, we examine the balance between BSA-driven acceleration and radiative energy losses, derive the corresponding maximum electron energies and steady-state spectra, and compare the resulting predictions with the observed radio spectral properties of representative cluster relics.

\section{Ballistic Surfing Acceleration: Physical Picture}

Ballistic surfing acceleration (BSA) \citep[][]{Stasiewicz2025} is a particle acceleration mechanism that follows directly from the electrodynamics of collisionless shocks.
The physical basis of BSA can be understood from the relativistic momentum equation for electrons with charge $e$, rest mass $m_e$, and velocity $\bm{v}$ in electromagnetic fields $\bm{E}$ and $\bm{B}$:
\begin{equation}
\frac{d\bm{p}}{dt} = e \left( \bm{E} + \frac{\bm{v}}{c} \times \bm{B} \right),
\end{equation}
where $\bm{p}=\gamma m_e\bm{v}$, $\gamma=(1-v^2/c^2)^{-1/2}$, and $c$ is the speed of light.
This implies that the kinetic energy gain of a particle is entirely due to the electric field,
\begin{equation}
\Delta \mathcal{E} = e \int \bm{E} \cdot \bm{v} dt,
\end{equation}
where $\mathcal{E} \equiv (m_e^2 c^4 + p^2 c^2)^{1/2} - m_e c^2$.

In general, there are three types of electric fields that determine particle dynamics and energization processes in the shock frame:
$\bm{E}_{\rm conv} = -\bm{V} \times \bm{B}/c$ --- the convection electric field;
$\hat{\bm{x}}E_S(x)$ --- the cross-shock electric field, maintained by the electron pressure gradient and directed along the shock normal $\hat{\bm{x}}$ aligned with the bulk velocity $ \bm{V}$;
and $\tilde{\bm{E}}(\bm{r},t)$ --- the wave electric field of arbitrary modes.
The convection electric field,  $E_{\rm conv} = (V_u/c) B_u \sin\zeta$ (where the subscript '$u$' denotes upstream values and $\zeta$ is the shock angle), remains constant across one-dimensional  structures in the shock reference frame.

The BSA framework was originally developed in the context of near-Earth collisionless shocks, where in-situ spacecraft measurements can resolve the shock ramp, wave electric fields, and particle orbits. In this space-plasma environment, thermal particles can first be energized by strong wave electric fields, $\tilde{E}\lesssim 100\,\mathrm{mV m^{-1}}$, observed at the Earth’s bow shock via the stochastic wave energization (SWE) mechanism \citep{Stasiewicz2023MN1,Stasiewicz2023MN2,Stasiewicz2023MN3, HaStasiewicz2026}. Once their gyroradius, $r_c$, exceeds the shock width ($r_c>\Delta$), the particles are no longer confined to the internal shock ramp and BSA becomes the dominant acceleration mechanism. The essential ingredients are then the large-scale convection electric field and the magnetic-field ramp, while pitch-angle scattering is required only to keep particles in the vicinity of the shock and to facilitate repeated gyrations around the magnetic ramp.

High-energy particles with gyroradii much larger than the width of the shock  (typically of the order of the thermal proton gyroradius) spend a negligible fraction of a gyroperiod inside the  ramp. Therefore, acceleration by the fields  $\bm{E}=\bm{E}_{\rm conv}+\hat{\bm{x}}E_S+\tilde{\bm{E}}$ inside the shock is negligible in comparison to acceleration by $\bm{E}_{\rm conv}$ outside the shock.

In this framework, the energization of high-energy particles is governed by their interaction with the large-scale electric field $\bm{E}_{\rm conv}$, rather than by stochastic scattering off magnetic turbulence and repeated shock crossings. This acceleration occurs during ``surfing'' outside the shock ramp, in a region where magnetic-field gradients and wave activity are weak. Consequently, the stochastic contribution to acceleration, $e\int \tilde{\mathbf{E}} \cdot \mathbf{v} \, dt$, averages to zero outside the shock, and $\nabla B$ drifts are negligible. BSA is thus entirely decoupled from processes occurring within the shock ramp and is conceptually distinct from shock drift acceleration (SDA) \citep[e.g.,][]{Armstrong1985, Decker1985, Katou2019}, which relies on $\nabla B$ drifts within the shock transition, applies primarily to low-energy particles ($r_c < \Delta$), and has been shown to be equivalent to quasi-adiabatic heating (QAH) \citep{Stasiewicz2023MN1}. In the SDA process, particles are energized as their gyrocenters execute $\nabla B$ drifts within the shock ramp. In the BSA process, particles gain energy by ``surfing'' within the large-scale convective electric field where $B \approx \mathrm{const}$. 

Both electrons and ions gain energy during upstream surfing ($q\bm{E}\cdot\bm{v} > 0$) and lose energy in the downstream region ($q\bm{E}\cdot\bm{v} < 0$), where $q = \pm e$. Since the upstream gyroradius is larger than the downstream gyroradius, $r_{cu}>r_{cd}$, the net energization arises from this asymmetry and can be estimated in magnitude as $\Delta \mathcal{E} \sim q(r_{cu} - r_{cd})E_\mathrm{conv}$.

The efficiency of BSA depends sensitively on the shock geometry. In quasi-perpendicular shocks, where the convection electric field is maximized, particles gain more energy during surfing in the low upstream magnetic field $B_u$ than they lose while gyrating in the compressed downstream field $B_d$, as clearly illustrated in Fig.~3 of \citet{Stasiewicz2025}. In contrast, for parallel shocks the convection electric field vanishes, and ballistic surfing acceleration becomes ineffective. This strong dependence on shock obliquity is a defining feature of the BSA mechanism and leads to acceleration efficiencies that vary continuously with shock angle.

Below, we apply the BSA model to galaxy cluster merger shocks, which are typically non-relativistic, weakly magnetized, and characterized by low Mach numbers. The energy gain per gyroperiod, averaged over an initially isotropic distribution function, is given by  \citep{Stasiewicz2025}:
\begin{equation}
\Delta \mathcal{E} \approx g(1-c_B^{-1})\frac{E_{\rm conv}}{B_u } \mathcal{E} ,
\label{DK}
\end{equation}
where $c_B = B_d/B_u$ is the magnetic compression ratio and $g \lesssim 1$ is a geometric factor dependent on the shock angle $\zeta$, which remains generally poorly constrained for distant shocks. For $g \approx 0.8$, the model yields a cosmic-ray spectral index of $s \approx -2.7$, in excellent agreement with observations \citep{Stasiewicz2025}.

Using Eq.~(\ref{DK}) for relativistic electrons with  kinetic energy $\mathcal{E} \approx  \gamma m_e c^2$, and gyrofrequency $\omega_c = eB/\gamma m_e c$, the energy gain rate per gyroperiod can be expressed in terms of the relativistic Lorentz factor as
\begin{equation}
\dot{\gamma}_{\rm BSA}
= \eta\,\frac{e E_{\rm conv}}{m_e c},
\qquad
\eta\equiv \frac{g}{4\pi}(c_B-c_B^{-1}).
\label{eq:gammadot_acc}
\end{equation}
For typical galaxy cluster shocks with $c_B \sim 2$--$3$, $\eta \sim 0.1$.

 Eq.~(\ref{eq:gammadot_acc}) represents the theoretical BSA acceleration capacity for particles with cyclotron orbits that intercept the shock front. However, such particles constitute only a minute fraction of the total electron population within galaxy cluster shocks. We therefore introduce a participation factor $\eta_{\rm BSA} \ll 1$ and define the effective acceleration rate, averaged over the entire source region, as:
\begin{equation}
\dot{\gamma}_{\rm acc} \approx \eta_{\rm BSA} \dot{\gamma}_{\rm BSA}.
\end{equation}
Here, $\eta_{\rm BSA}$ represents a macroscopic participation factor that accounts for the fact that shocks occupy only a very small volume within galaxy clusters. Consequently, the particles currently associated with shock-front acceleration represent only a small and generally unknown fraction of the total population of radiating particles accelerated at earlier times.

Several important features of the acceleration rate follow immediately from this formulation. First, the acceleration rate is independent of particle energy, implying linear energy growth in time in the absence of losses. Second, the acceleration efficiency depends explicitly on the shock obliquity through the geometric factor $g$ and $E_{\rm conv}$. It vanishes in the limit of parallel shocks, where the motional electric field disappears, and also in the absence of shocks, when $c_B=1$. Finally, the acceleration rate depends only on macroscopic shock parameters and does not invoke particle diffusion coefficients or turbulent scattering rates.

This description is therefore well suited for application to galaxy cluster merger shocks, where large-scale electrodynamic fields are robustly present while the microphysics of particle scattering remains uncertain. In the following sections, this acceleration rate is combined with radiative loss processes to determine the maximum electron energy and the resulting steady-state electron spectrum.

\section{Galaxy Cluster Merger Shock Environment}
Galaxy cluster merger shocks propagate through the intracluster medium, a hot, dilute plasma characterized by temperatures of several keV \citep[e.g.,][]{Sarazin1986,Voit2005} and particle densities of order $10^{-4}$--$10^{-3}\,\mathrm{cm^{-3}}$ \citep[e.g.,][]{Eckert2012}. Observations indicate that typical shock velocities range from a few hundred to approximately $10^3\,\mathrm{km\,s^{-1}}$ \citep[e.g.,][]{Markevitch2007}, while magnetic field strengths in the cluster outskirts are commonly estimated to be of order $\mu$G or lower \citep[e.g.,][]{Murgia2004,Bruggen2012}. These conditions correspond to non-relativistic shocks with small Mach numbers \citep[e.g.,][]{Miniati2001, Ryu2003, Vazza2009, Ha2018}.

Compared with the Earth's bow shock, where BSA was originally developed using in situ spacecraft measurements, galaxy cluster merger shocks occur in a distinct plasma environment. The Earth’s bow shock is embedded in the relatively dense, magnetized solar wind with kinetic scales directly resolvable by spacecraft. Cluster shocks, by contrast, propagate through a much hotter, dilute intracluster medium characterized by $\mu\text{G}$-level magnetic fields, low sonic Mach numbers, high plasma beta, and Mpc-scale surfaces evolving over cosmological timescales. These differences imply that injection-scale physics cannot be observed directly in clusters; consequently, the effective BSA efficiency must be interpreted as a macroscopic, source-averaged quantity. Nevertheless, the essential electrodynamic ingredients required for BSA—a magnetic-field ramp, a motional electric field, and favorable quasi-perpendicular geometry—are inherently present in both environments. This application to radio relics should therefore be viewed as an extrapolation of the local BSA mechanism to the large-scale conditions of cluster merger shocks, where its viability can be tested indirectly through synchrotron emission.

For relativistic electrons in the intracluster medium, radiative energy losses play a central role in determining the shape and extent of the particle energy distribution. The dominant loss processes are synchrotron radiation in the cluster magnetic field and inverse-Compton scattering off cosmic microwave background (CMB) photons \citep[e.g.,][]{Webb1984, Sarazin1999}. Both processes arise from interactions between relativistic electrons and ambient electromagnetic fields and lead to continuous energy losses that scale quadratically with the electron Lorentz factor \citep[e.g.,][]{Rybicki1979}.
The total radiative energy loss rate can be written as
\begin{equation}
\dot{\gamma}_{\rm loss}
= - \frac{4}{3}\,\frac{\sigma_T }{m_e c}\,
\left( U_B + U_{\rm CMB} \right)\,\gamma^2,
\label{eq:gammadot_loss}
\end{equation}
where $\sigma_T$ is the Thomson cross section. The quantity $U_B$ denotes the magnetic field energy density associated with synchrotron losses, while $U_{\rm CMB}$ represents the energy density of the cosmic microwave background radiation responsible for inverse-Compton cooling.
The magnetic energy density is given by
\begin{equation}
U_B = \frac{B^2}{8\pi},
\end{equation}
where $B$ is the local magnetic field strength, while the CMB energy density evolves with redshift as
\begin{equation}
U_{\rm CMB} = U_{\rm CMB,0}\,(1+z)^4,
\end{equation}
with $U_{\rm CMB,0}$ denoting the present-day CMB energy density.

Eq.~(\ref{eq:gammadot_loss}) follows from classical electrodynamics in the Thomson regime and applies as long as the electron energies relevant for radio synchrotron emission satisfy $\gamma h\nu_{\rm CMB} \ll m_e c^2$, a condition well satisfied for cluster radio relic electrons. The quadratic dependence on $\gamma$ reflects the fact that both synchrotron and inverse-Compton losses scale with the square of the electron energy.

In galaxy clusters, the relative importance of synchrotron and inverse-Compton losses is determined by the ratio $U_B/U_{\rm CMB}$. For typical magnetic field strengths of  $0.1 - 1~\mu$G in cluster outskirts, the magnetic energy density is often comparable to or smaller than the CMB energy density, particularly at moderate redshifts \citep[][]{Sarazin1999}. As a result, inverse-Compton cooling frequently dominates the total radiative losses of relativistic electrons in radio relics.

In principle, radiative losses depend on the total electromagnetic acceleration experienced by a charged particle. One may therefore ask whether the convection electric field associated with the shock contributes an additional radiative loss term. For cluster merger shocks, however, the convection electric field satisfies $E_{\rm conv} = (V_u/c)B_u$, where the upstream velocity $V_u$ is much smaller than the speed of light. As a result, the radiative contribution associated with $E_{\rm conv}$ is suppressed relative to the magnetic-field-driven transverse acceleration by a factor of order $(V_u/c)^2$. The standard synchrotron loss expression therefore provides an excellent approximation under intracluster conditions.

Any viable acceleration mechanism operating in cluster merger shocks must therefore compete effectively with these radiative losses in order to sustain electrons at Lorentz factors $\gamma \sim 10^4$--$10^5$ required for observable radio synchrotron emission \citep[e.g.,][]{Hoeft2007, Siemieniec-Ozieblo2011, Pinzke2013}. For representative cluster conditions, the shock-transition scale may be taken as a few thermal ion gyroradii, $\Delta \sim {\rm a~few}\, r_{c,i,\rm th}$ \citep[e.g.,][]{Ha2021}. 
Using $T_i \sim {\rm a~few}\,{\rm keV}$ and $B \sim 0.1$--$1\,\mu{\rm G}$, this gives $\Delta \sim 10^{10}$--$10^{11}\,{\rm cm}$. 
By contrast, electrons responsible for observable radio relic emission, with $\gamma \sim 10^4$--$10^5$, have gyroradii $r_{c,e} \sim 10^{13}$--$10^{15}\,{\rm cm}$, implying $r_{c,e}/\Delta \sim 10^2$--$10^4$. 
Thus, the condition $r_{c,e} \gg \Delta$ is well satisfied in the energy range relevant for radio synchrotron emission, placing the electron dynamics in a regime where ballistic transport dominates, while non-adiabatic or stochastic effects \citep[e.g.,][]{Katou2019} are expected to be important only in the transitional regime $r_c \sim \Delta$. In the following section, we combine the acceleration rate derived for BSA with the loss rate given above to determine the maximum electron energy attainable in cluster shocks.

\section{Maximum Electron Energy from Acceleration--Loss Balance}

The maximum energy attainable by electrons accelerated via ballistic surfing acceleration
is determined by the competition between systematic energy gain and radiative losses.
In the steady-state limit, the highest electron energy is reached when the acceleration
timescale equals the radiative cooling timescale, corresponding to a balance between
energy gain and loss rates.

This condition provides a direct and physically transparent estimate of the maximum
electron Lorentz factor attainable in galaxy cluster merger shocks.
Importantly, the resulting expression depends only on macroscopic shock properties—such as
the upstream flow velocity, magnetic field strength, and redshift—together with an
effective efficiency parameter describing the ballistic surfing process.
No assumptions regarding particle diffusion coefficients or turbulent scattering rates
are required.

\begin{figure*}
    \centering
    \includegraphics[width=\textwidth]{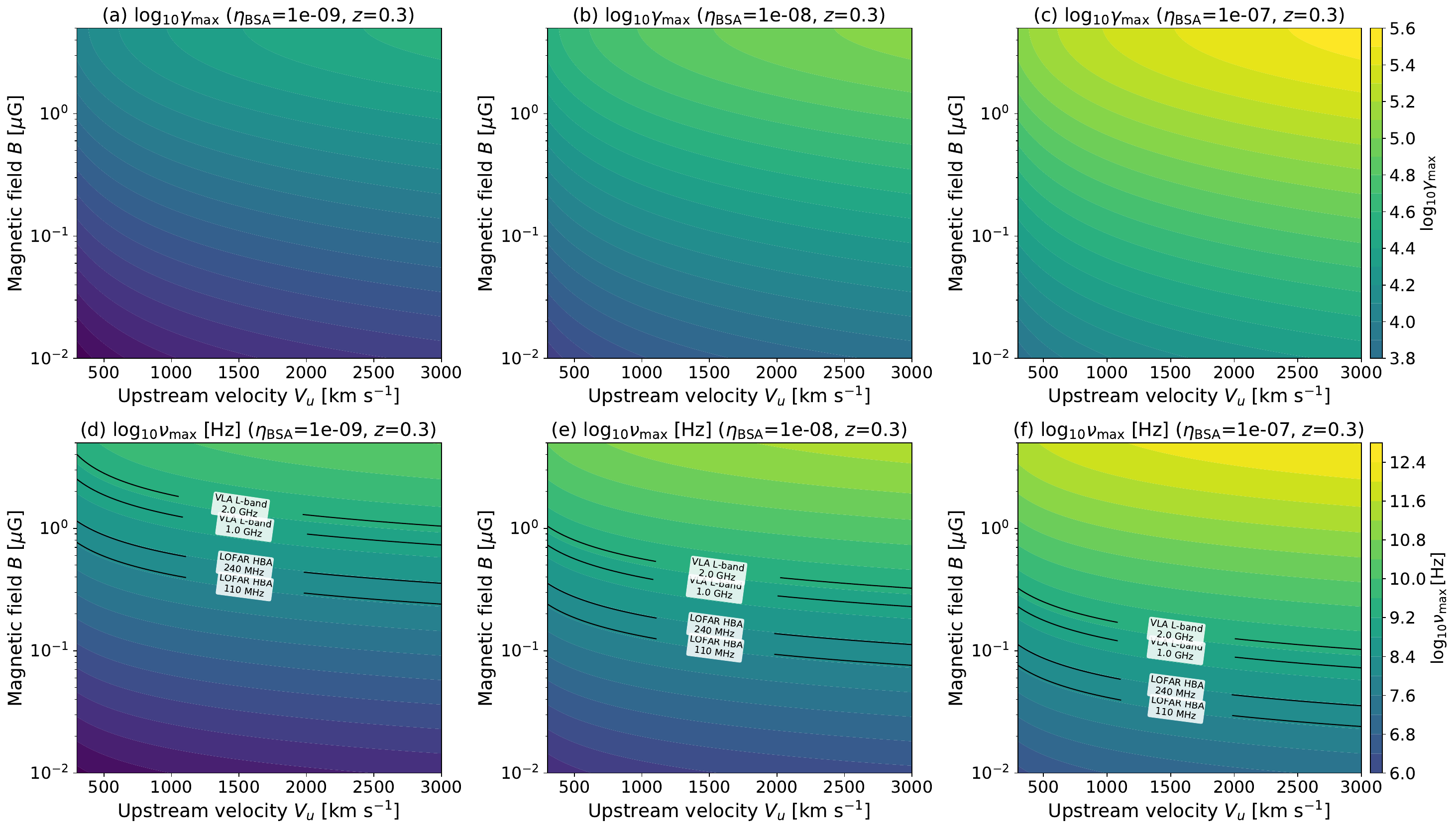}
    \caption{
    Mapping between ballistic surfing acceleration and observable radio emission in galaxy cluster merger shocks.
    The upper panels show the maximum electron Lorentz factor $\gamma_{\max}$, determined by the balance between BSA and radiative losses, as a function of upstream velocity $V_u$ and magnetic field strength $B$ for two representative values of $\eta_{\rm BSA}$.
    The lower panels show the corresponding maximum synchrotron frequency $\nu_{\max}$ inferred from $\gamma_{\max}$, with the characteristic observing bands of LOFAR (HBA) \citep[][]{vanHaarlem2013} and the VLA (L-band) \citep[][]{Perley2009} indicated.
    }
    \label{fig:f1}
\end{figure*}

\subsection{Acceleration--Cooling Balance and Maximum Lorentz Factor}

The balance condition can be expressed as
\begin{equation}
\dot{\gamma}_{\rm acc} + \dot{\gamma}_{\rm loss} = 0,
\end{equation}
where $\dot{\gamma}_{\rm acc}$ is the acceleration rate associated with ballistic surfing
and $\dot{\gamma}_{\rm loss}$ is the combined synchrotron and inverse-Compton loss rate.
Substituting Eqs.~(\ref{eq:gammadot_acc}) and (\ref{eq:gammadot_loss}), the maximum
electron Lorentz factor is given by
\begin{equation}
\gamma_{\max} \approx
\left[
\frac{3\,\eta\, \eta_{\rm BSA}\, e E_{\rm conv}}
{4\,\sigma_T \,(U_B + U_{\rm CMB})}\,
\right ]^{1/2}.
\label{eq:gammamax}
\end{equation}
The maximum electron energy can be estimated by comparing the acceleration timescale, $t_{\rm acc}=\gamma/\dot{\gamma}_{\rm acc}$, with the radiative cooling timescale, $t_{\rm cool}=\gamma/|\dot{\gamma}_{\rm loss}|$. 
The condition $t_{\rm acc} = t_{\rm cool}$ defines the maximum Lorentz factor $\gamma_{\max}$, indicating that the attainable electron energy is set by the competition between coherent electrodynamic acceleration and radiative losses.

\subsection{From Maximum Electron Energy to Observable Radio Frequencies}

While Eq.~(\ref{eq:gammamax}) characterizes the efficiency of ballistic surfing
acceleration in terms of particle energy, radio relics are observed through their
synchrotron emission at specific frequencies.
A direct comparison with observations therefore requires translating the maximum
electron Lorentz factor into the corresponding characteristic synchrotron frequency.

For relativistic electrons emitting in a magnetic field $B$, the characteristic
synchrotron frequency is given by
\begin{equation}
\nu_c \approx \frac{3 e B}{4\pi m_e c}\,\gamma^2.
\end{equation}
This relation establishes a direct mapping between the particle energy space and
observable radio frequencies.

In Fig.~\ref{fig:f1}, we evaluate Eq.~(\ref{eq:gammamax}) over a broad range of representative cluster merger shock parameters and show both the resulting maximum electron Lorentz factor $\gamma_{\max}$ and the corresponding maximum synchrotron frequency $\nu_{\max}$.
In the cluster outskirts, radiative losses of relativistic electrons are typically dominated by inverse-Compton (IC) scattering off the cosmic microwave background (CMB), because $U_{\rm CMB}$ generally exceeds the magnetic energy density $U_B=B^2/8\pi$ for $\mu$G-level fields. In this regime, the cooling rate is well approximated by $\dot{\gamma}_{\rm loss}\propto -U_{\rm CMB}\gamma^2$, and depends only weakly on $B$.
In contrast, BSA yields an approximately energy-independent acceleration rate $\dot{\gamma}_{\rm acc}\propto \eta_{\rm BSA} V_u B$, set by the large-scale convective electric field upstream of the shock.

The parameter $\eta_{\rm BSA}$ controls the effective acceleration rate, accounting for both the fraction of electrons that participate in coherent surfing and the degree of phase coherence in their trajectories. It should not be interpreted as an arbitrary normalization parameter, but as a source-averaged participation and coherence factor that reduces the ideal single-particle BSA rate to the effective rate realized in an extended relic region. The maps shown in Fig.~\ref{fig:f1} do not represent a spectral fit; rather, they illustrate how the balance between acceleration and synchrotron losses depends on representative values of $\eta_{\rm BSA}$, $B$, and $V_u$. A useful heuristic range for $\eta_{\rm BSA}$ is obtained by requiring typical cluster shocks with $V_u \approx 10^3\text{ km s}^{-1}$ and $B \approx 0.1\text{--}1\text{ }\mu\text{G}$ to accelerate electrons to Lorentz factors of $\gamma_{\max} \approx 10^4\text{--}10^5$, as required for observable radio relic emission. Increasing either $B$ or $\eta_{\rm BSA}$ shifts this balance toward higher energies, expanding the region of parameter space where $\gamma_{\max}$ and the cutoff frequency $\nu_{\max}$ are sufficient to account for observed radio relics.

\section{Steady-State Electron Spectrum and Synchrotron Emission}

\subsection{Steady-State Electron Spectrum}
\label{sec:transport}

Beyond estimating the maximum electron energy, a meaningful comparison with radio relic
observations requires modeling the steady-state energy distribution of relativistic
electrons established by the interplay between acceleration, injection, and radiative
losses. To this end, we consider a simplified transport equation describing the evolution
of the electron distribution function $N(\gamma)$ in energy space.

\begin{figure}
    \centering
    \includegraphics[width=\columnwidth]{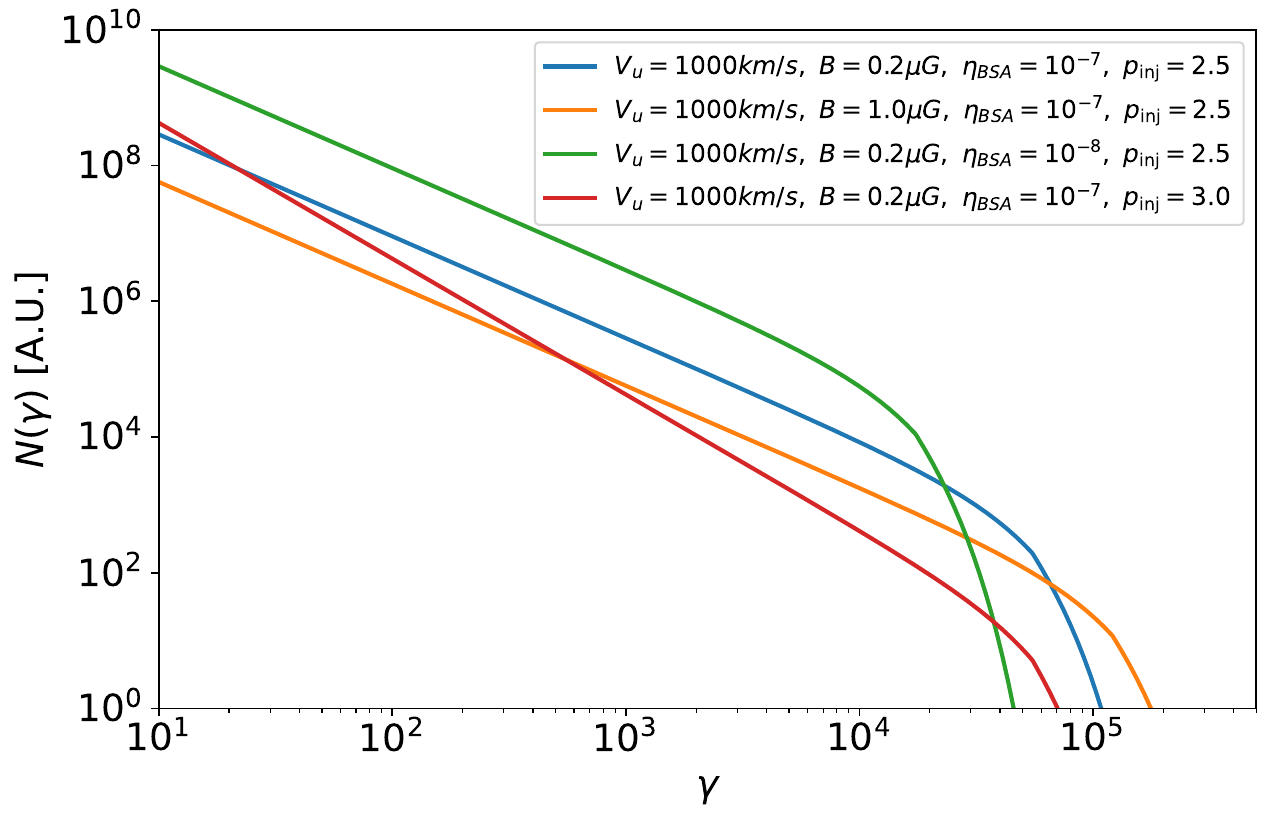}
    \caption{
    Steady-state electron energy distributions $N(\gamma)$ produced by ballistic surfing acceleration in galaxy cluster merger shocks. Different curves illustrate the dependence on magnetic field strength, BSA efficiency, and injection spectral index, while the high-energy steepening near $\gamma_{\max}$ reflects radiative cooling.
    }
    \label{fig:f2}
\end{figure}

The electron population is governed by the kinetic equation
\begin{equation}
\frac{d}{d\gamma}
\left[
(\dot{\gamma}_{\rm acc} + \dot{\gamma}_{\rm loss})\,N(\gamma)
\right]
+ \frac{N(\gamma)}{t_{\rm esc}}
= Q(\gamma),
\label{eq:transport}
\end{equation}
In solving Eq.~(\ref{eq:transport}), we adopt a power-law injection term,
$Q(\gamma)=Q_0\,\gamma^{-q_{\rm inj}}$ for $\gamma\ge \gamma_{\rm inj,min}$, where
$\gamma_{\rm inj,min}$ represents the minimum Lorentz factor of seed electrons.
For a meaningful comparison across different shock and magnetic-field parameters,
the normalization $Q_0$ is not kept fixed; instead, it is determined by requiring a
fixed \emph{injected energy rate} into relativistic electrons,
\begin{equation}
\dot{\mathcal{E}}_{\rm inj} \propto \int_{\gamma_{\rm inj,min}}^{\gamma_{\max}} \gamma\,Q(\gamma)\,d\gamma
= \mathrm{const.}
\label{eq:Einj_norm}
\end{equation}
This choice ensures that differences among the steady-state solutions arise from the
competition between ballistic surfing acceleration and radiative losses (through the
parameter-dependent $\gamma_{\max}$), rather than from an arbitrary change in the overall
injection power.
Solving Eq.~(\ref{eq:transport}) in the cooling-dominated regime then yields the
steady-state electron energy distributions shown in Fig.~\ref{fig:f2}. We observed that the trend shown in Fig.~\ref{fig:f2} is fully consistent with the $\gamma_{\max}(V_u,B)$ maps in Fig.~\ref{fig:f1}.

We also note that particle escape does not control the high-energy spectrum in the relic environment. The characteristic radiative cooling time for electrons emitting at radio frequencies ($\gamma \sim 10^{4}$--$10^{5}$) is typically much shorter than the macroscopic advection or leakage time across the acceleration region, which is of the order $t_{\rm esc} \sim L/V_u$ for a region of size $L$. Given the large spatial scales of cluster merger shocks and relic widths ($L \sim 10^2\,\mathrm{kpc}$) and shock speeds of $V_u \sim 10^{3}\,\mathrm{km\,s^{-1}}$, the escape time $t_{\rm esc} \sim 10^2\,\mathrm{Myr}$ is significantly longer than both the cooling time $t_{\rm cool} \sim 10\,\mathrm{Myr}$ and the effective acceleration time $t_{\rm acc} \sim 10\,\mathrm{Myr}$ within the energy range of interest. This estimate assumes an advection-dominated escape timescale and provides an effective description of particle transport in the downstream region. While this treatment neglects multi-dimensional geometry and potential contributions from diffusive transport in extended halo structures, such effects are not expected to qualitatively modify the high-energy spectral shape, which remains primarily governed by the competition between acceleration and radiative losses.

\subsection{Synchrotron Emission from BSA-Accelerated Electrons}

The synchrotron spectra produced by BSA-accelerated electrons exhibit a characteristic curvature that arises naturally from the competition between energy-independent acceleration and quadratic radiative losses.
Using the steady-state electron distributions derived in Section~\ref{sec:transport},
we compute the resulting synchrotron emission by forward modeling the radiative output
of relativistic electrons in the intracluster magnetic field.
For a given isotropic electron distribution $N(\gamma)$, the synchrotron emissivity
per unit frequency in the emitting frame is written as
\begin{equation}
j_{\nu,\mathrm{em}}
=
\int_{\gamma_{\min}}^{\gamma_{\max}}
N(\gamma)\,P_{\rm syn}(\nu_{\mathrm{em}},\gamma)\,d\gamma ,
\label{eq:jsyn}
\end{equation}
where $P_{\rm syn}(\nu,\gamma)$ is the single-electron synchrotron power per unit frequency
and $\nu_{\mathrm{em}}=(1+z)\nu_{\mathrm{obs}}$ accounts for cosmological redshift.
For an isotropic pitch-angle distribution, the single-electron synchrotron power is
\begin{equation}
P_{\rm syn}(\nu,\gamma)
=
\frac{\sqrt{3}\,e^3 B}{m_e c^2}\,
F\!\left(\frac{\nu}{\nu_c}\right),
\qquad
\nu_c=\frac{3eB}{4\pi m_e c}\,\gamma^2 ,
\label{eq:psyn}
\end{equation}
where $F(x)=x\int_x^\infty K_{5/3}(\xi)\,d\xi$ is the standard synchrotron kernel
and $K_{5/3}$ denotes the modified Bessel function of the second kind.
This formulation maps the electron energy distribution directly into an observable
radio spectrum without invoking a phenomenological power-law prescription.

\begin{figure}
    \centering
    \includegraphics[width=\columnwidth]{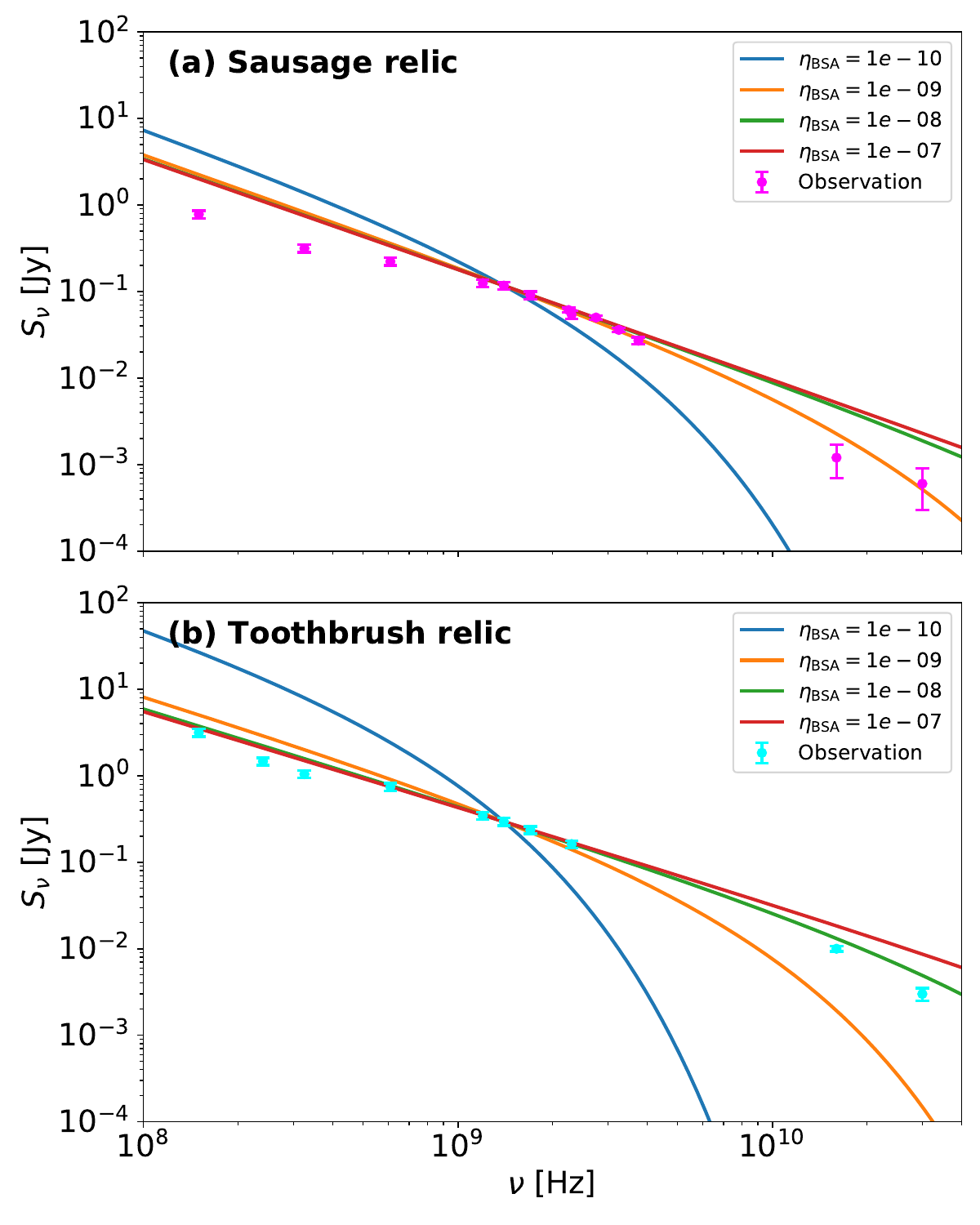}
    \caption{
    Synchrotron spectra produced by BSA–accelerated electrons in galaxy cluster shocks, compared with integrated radio relic observations. The upper and lower panels show the Sausage and Toothbrush relics, respectively. Solid curves denote model spectra for different values of the BSA efficiency parameter $\eta_{\rm BSA}$. Points with error bars indicate observed integrated flux densities from \citet{Stroe2016}. The uncertainties are typically of order $\sim 10\%$ and are mostly comparable to the marker size on the logarithmic scale.
    }
    \label{fig:f3}
\end{figure}

To relate the BSA-driven acceleration spectrum to the emitting electron
population, we assume that BSA operates primarily
in the upstream region and within the shock transition layer, where electrons
are energized up to a maximum Lorentz factor $\gamma_{\max}$ set by the balance
between coherent acceleration and radiative losses. Upon crossing the shock,
this accelerated population is injected into the downstream flow with the
cutoff distribution described below. Subsequent evolution is assumed to be
governed by radiative cooling and advection, while any additional downstream
acceleration is neglected.

Under these assumptions, we compute the electron distribution in the emission
region by solving the steady-state transport equation including radiative cooling
and particle escape,
\begin{equation}
\frac{d}{d\gamma}
\left(
\dot{\gamma}_{\rm loss}\,N(\gamma)
\right)
+\frac{N(\gamma)}{t_{\rm esc}}
\approx
Q(\gamma),
\label{eq:ss_cooling}
\end{equation}
where $Q(\gamma)$ denotes the injected electron spectrum,
$t_{\rm esc}$ is the characteristic escape (or advection) timescale from the
emission region, and the radiative energy loss rate is
$\dot{\gamma}_{\rm loss}=-a_{\rm loss}\gamma^2$, with
\begin{equation}
a_{\rm loss}
=
\frac{4}{3}\,
\frac{\sigma_T c}{m_e c^2}
\left(U_B+U_{\rm CMB}\right).
\end{equation}
The resulting solution can be written as
\begin{equation}
N(\gamma)
\approx
\frac{1}{|\dot{\gamma}_{\rm loss}|}
\int_\gamma^{\infty}
Q(\gamma')
\exp\left[
-\int_\gamma^{\gamma'}
\frac{d\tilde{\gamma}}
{|\dot{\gamma}_{\rm loss}(\tilde{\gamma})|\,t_{\rm esc}}
\right]
d\gamma',
\label{eq:Ngamma}
\end{equation}
which describes an electron population shaped by the combined effects of
radiative cooling and finite residence time downstream of the shock.
In Eq. (\ref{eq:Ngamma}), the injected electron spectrum $Q(\gamma)$ is modeled as a power-law with an exponential cutoff,
\begin{equation}
Q(\gamma) \propto \gamma^{-p_{\rm inj}}
\exp\left(-\frac{\gamma}{\gamma_{\max}}\right),
\end{equation}
where $p_{\rm inj}$ characterizes a pre-existing population of suprathermal electrons in the intracluster medium. 
In this framework, BSA acts primarily as a re-acceleration mechanism, boosting these seed electrons to higher energies.
The maximum Lorentz factor $\gamma_{\max}$ is determined by the balance between coherent electrodynamic acceleration and radiative losses. 
As a result, BSA enters the synchrotron model mainly by setting the high-energy cutoff of the electron population.
While the resulting radio spectra are not uniquely determined by the BSA mechanism alone, they demonstrate that the maximum electron energies attainable through BSA are consistent with the observed high-frequency steepening of radio relics. 
This highlights the role of BSA in establishing the accessible energy range of relativistic electrons under cluster shock conditions.

The observed flux-density spectrum is obtained from the emitted spectrum by
accounting for cosmological redshift.
For a direct comparison of spectral shapes, each model spectrum is normalized to match the observed flux density
at a reference frequency $\nu_0=1.4\,\mathrm{GHz}$,
\begin{equation}
S_{\nu}^{\rm model}(\nu)
\approx
j_{\nu}(\nu)\,
\frac{S_{\nu,\mathrm{obs}}(\nu_0)}{j_{\nu}(\nu_0)} .
\label{eq:spec_norm}
\end{equation}
This procedure isolates the spectral curvature and high-frequency rollover
introduced by ballistic surfing acceleration and radiative cooling,
without introducing additional free parameters associated with the emitting volume
or source distance.

Fig.~\ref{fig:f3} shows the synchrotron spectra produced by BSA-accelerated electrons for a range of values of the effective efficiency parameter $\eta_{\rm BSA}$, compared with the integrated radio spectra of the Sausage and Toothbrush relics.
The error bars indicate the reported flux-density uncertainties from \citet{Stroe2016}, which are typically of order $\sim 10\%$ and are mostly comparable to the marker size on the logarithmic scale. The comparison should therefore be interpreted as a test of the overall spectral curvature and high-frequency steepening rather than as a formal multi-parameter fit.
The resulting spectra exhibit a characteristic curvature and high-frequency steepening, which arise naturally from the competition between energy-independent acceleration and quadratic radiative losses.
The comparison with observations shows that only a narrow range of $\eta_{\rm BSA}\sim10^{-9}-10^{-8}$ is consistent with the data. Larger values lead to spectra that are too hard due to excessively large $\gamma_{\max}$, while smaller values fail to produce electrons energetic enough to account for the observed GHz-frequency emission.
This behavior indicates that the observed spectral curvature directly constrains the effective acceleration rate, implying that $\eta_{\rm BSA}$ is not a free tuning parameter but an observable proxy for the macroscopic efficiency of coherent electrodynamic energization. The radio relic spectra therefore provide a direct probe of the balance between acceleration and radiative losses in cluster merger shocks.

The small value of $\eta_{\rm BSA}$ reflects the intrinsically limited fraction of electrons that can participate in coherent surfing, rather than a weakness of the underlying acceleration mechanism. 
In realistic cluster merger shocks, only particles whose trajectories satisfy the geometric condition for ballistic surfing contribute to sustained energization.
As a result, $\eta_{\rm BSA}$ encodes the combined effects of finite upstream residence times, orbit decoherence, and geometric constraints, providing a measure of the effective macroscopic efficiency of coherent electrodynamic acceleration. 
The inferred values therefore arise naturally from the transport properties of the system. 

Despite this strong reduction, the inferred efficiency remains sufficient to
accelerate electrons to Lorentz factors $\gamma\sim10^{4}$--$10^{5}$ in the
intracluster medium, owing to the energy-independent acceleration rate of BSA and the dominance of inverse-Compton cooling. The resulting spectra are reproduced in a physically self-consistent manner, indicating that BSA can provide a viable channel for generating relativistic electrons in cluster
merger shocks. 
More broadly, these results suggest that coherent electrodynamic processes may contribute to shaping the nonthermal electron population in radio relics. Whether such processes operate independently or in combination with other proposed acceleration mechanisms remains an open question that warrants further investigation.

\subsection{Geometry and observational implications}

The present framework is primarily intended to reproduce the integrated synchrotron spectrum of radio relics. Several spatially resolved observables—including polarization, surface-brightness profiles, spectral-index gradients, and line-of-sight projection effects—therefore deserve additional discussion in the context of the BSA scenario.

An important observational property of many radio relics is their high degree of linear polarization, generally interpreted as evidence for ordered magnetic fields associated with merger shocks \citep[e.g.,][]{vanWeeren2010,vanWeeren2019}. High-resolution observations reveal substantial polarization fractions and downstream depolarization trends, likely related to magnetic-field geometry and line-of-sight effects \citep[e.g.,][]{DiGennaro2021}. In this framework, BSA is regarded as an electron energization mechanism rather than a driver of large-scale magnetic topology. Polarization properties are thus controlled by shock geometry, magnetic-field compression, downstream magnetic-field ordering, and line-of-sight integration, while BSA constrains the energy distribution and high-energy cutoff of the radiating electrons. Since BSA is most efficient in quasi-perpendicular shocks, where the motional electric field reaches its maximum, it is compatible with the ordered downstream magnetic configurations commonly invoked to explain high polarization fractions.

Geometrical effects are also critical for interpreting the morphology and spectral-index structure of radio relics. Observed emission depends not only on the local acceleration mechanism but also on the curvature of the shock surface, the spatial distribution of the magnetic field, downstream advection, radiative cooling, and projection along the line of sight. BSA determines the local electron energization and maximum attainable energy, whereas the observed surface-brightness distribution results from subsequent transport and radiative evolution. Line-of-sight integration through a curved shock surface can broaden the apparent relic width and modify observed spectral gradients, even if the underlying acceleration process is localized at the shock. Such projection effects are expected to affect observed morphology independently of whether the electron energization is produced by BSA or by a diffusive process.

Transverse surface-brightness and spectral-index profiles across radio relics provide another useful diagnostic. If electrons are energized locally at the shock and subsequently advected downstream, the width of the radio-emitting region is controlled by the competition between downstream advection and radiative cooling. Higher-frequency emission arises from a narrower region closer to the shock front because those electrons have shorter cooling times. This qualitative behavior is similar to that expected in conventional models, where the downstream profile is shaped by synchrotron and inverse-Compton losses, magnetic-field evolution, and advection \citep[e.g.,][]{Donnert2016}. In the BSA scenario, the primary difference is that the high-energy cutoff of the injected electron population is set by coherent electrodynamic acceleration rather than diffusive acceleration. Thus, BSA primarily affects the maximum emitting frequency and spectral steepening across the relic, while the detailed surface-brightness profile remains dependent on downstream transport, magnetic-field structure, shock curvature, and line-of-sight projection.

Pre-accelerated or fossil electron populations may also influence the interpretation of BSA-powered spectra. In this model, the injected electron spectrum is treated phenomenologically via $Q(\gamma)$, where $p_{\rm inj}$ characterizes the slope of the seed suprathermal population. If fossil electrons—supplied by previous AGN activity or earlier episodes of turbulent acceleration—are present near the relic, BSA need not accelerate particles directly from the thermal pool. Instead, it can act as a re-acceleration mechanism, raising the existing suprathermal population to the Lorentz factors required for radio synchrotron emission. Such a seed population reduces the severity of the injection barrier associated with the condition $r_{c,e}>\Delta$, potentially modifying both the low-energy normalization and the apparent injection slope. In this sense, BSA is not mutually exclusive with fossil-electron scenarios; pre-accelerated electrons may provide favorable initial conditions for coherent surfing acceleration. Relics associated with nearby AGN activity or remnant plasma provide useful targets for testing whether $p_{\rm inj}$ and $\eta_{\rm BSA}$ depend on the availability of fossil seed electrons.

A fully quantitative prediction of polarization maps, cross-sectional brightness profiles, and spatially resolved spectral-index distributions requires multi-scale modeling in which the BSA prescription is coupled to multidimensional cluster-shock dynamics, magnetic-field evolution, downstream electron transport, and synthetic synchrotron radiative-transfer calculations. Such modeling, potentially involving idealized merger-shock simulations or cosmological magnetohydrodynamic simulations, is beyond the scope of this paper. Nevertheless, the considerations above indicate that the observed polarization and morphological properties of radio relics are compatible with the BSA scenario and provide promising diagnostics for future studies.

\subsection{Injection Constraint and Energetic Viability of Ballistic Surfing Acceleration}

A useful measure of the global impact of BSA
is obtained by linking particle-level energization to the macroscopic
energy budget of the shock.
The instantaneous energy gain of a single electron is 
$\dot{\mathcal{E}}(\gamma)=m_ec^2\dot{\gamma}_{\rm acc}$,
where $\dot{\gamma}_{\rm acc} \approx \eta_{\rm BSA}\dot{\gamma}_{\rm BSA}$.
For a relativistic electron population described by the distribution
$N(\gamma)$ within the energization region, the volumetric power
deposited into electrons can be written as

\begin{equation}
\dot{U}_e^{\rm (BSA)}
=
\int_{\gamma_{\min}}^{\gamma_{\max}}
N(\gamma)\,m_ec^2\dot{\gamma}_{\rm acc}\,d\gamma
=
m_ec^2\dot{\gamma}_{\rm acc}\,n_{e,{\rm inj}},
\end{equation}
where
$n_{e,{\rm inj}}\equiv\int N(\gamma)\,d\gamma$
is the number density of accelerated electrons.
The associated energy flux can be estimated as
$F_e^{\rm (BSA)}\sim\dot{U}_e^{\rm (BSA)}L$,
where $L$ characterizes the effective residence scale.
Comparing this with the shock kinetic energy flux
$F_{\rm kin}=\frac{1}{2}\rho V_u^3$
yields a global acceleration efficiency
\begin{equation}
\xi_e
\equiv
\frac{F_e^{\rm (BSA)}}{F_{\rm kin}}
\sim
\frac{
m_ec^2\,\eta_{\rm BSA}\dot{\gamma}_{\rm acc}\,
n_{e,{\rm inj}}\,L
}{
\frac{1}{2}\rho V_u^3
},
\label{efficiency}
\end{equation}
providing a direct connection between coherent electrodynamic
energization and the shock energy reservoir.

The parameter $n_{e,{\rm inj}}$ can be physically interpreted
through an injection fraction,
$f_{e, {\rm inj}}\equiv n_{e,{\rm inj}}/n_e$.
BSA requires that electrons possess
gyroradii exceeding the thickness of the shock transition layer.
In collisionless shocks, the ramp width is typically comparable
to a few thermal proton gyroradii, implying that the majority
of thermal electrons remain magnetically tied to the shock and
do not undergo coherent surfing.
Consequently, only electrons belonging to the suprathermal tail
of the distribution satisfy the geometric condition
$r_{c,e} > \Delta$ and can experience sustained
electrodynamic acceleration.
This requirement naturally introduces an injection barrier,
indicating that BSA is intrinsically injection-limited.
The global electron energy flux therefore scales with the product
$f_{\rm e,inj}\eta_{\rm BSA}$, reflecting both the fraction of
electrons that participate in surfing and the efficiency with
which those electrons are energized.

For representative cluster merger shock parameters
($V_u\sim10^3\,{\rm km\,s^{-1}}$,
$n\sim10^{-4}\,{\rm cm^{-3}}$,
$B\sim1\,\mu{\rm G}$,
and $L\sim100\,{\rm kpc}$),
the resulting kinetic energy flux is
$F_{\rm kin}\sim10^{-4}\,{\rm erg\,cm^{-2}\,s^{-1}}$.
For a characteristic shock surface of order $\sim {\rm Mpc}^2$, this implies a total kinetic power of $\sim 10^{45}~{\rm erg~s^{-1}}$
Using these parameters, Eq.~(\ref{efficiency}) can be written in a convenient normalized form as
\begin{equation}
\xi_e \approx
2.8 \times 10^{-4}
\left(\frac{f_{e,{\rm inj}}}{10^{-9}}\right)
\left(\frac{\eta_{\rm BSA}}{10^{-9}}\right).
\end{equation}
Adopting a physically motivated, geometrically constrained
injection fraction from the thermal Maxwellian electron
distribution, $f_{\rm e,inj}\sim10^{-9}$,
together with $\eta_{\rm BSA}\sim10^{-9}$--$10^{-8}$
yields a global acceleration efficiency
$\xi_e\sim10^{-4}$--$10^{-3}$,
comfortably within the energy budget of cluster merger shocks.
Such a small injection fraction is naturally expected if only
electrons with gyroradii exceeding the shock thickness can
undergo sustained surfing.
These considerations indicate that the small electrodynamic
efficiency inferred from radio spectra is fully compatible with
macroscopic energetics once the injection constraint is taken
into account.

\subsection{Consistency with Inverse-Compton Constraints}

The same population of relativistic electrons responsible for the synchrotron emission in radio relics inevitably upscatters CMB photons via inverse-Compton  scattering. Any viable acceleration scenario must, therefore, remain consistent with existing X-ray and $\gamma$-ray constraints on non-thermal emission from galaxy clusters.

In the Thomson regime relevant for relic electrons with Lorentz factors $\gamma \sim 10^{4}$--$10^{5}$, the ratio of IC to synchrotron power is approximately

\begin{equation}
\frac{P_{\rm IC}}{P_{\rm syn}} \approx \frac{U_{\rm CMB}}{U_B}
= \left(\frac{B_{\rm CMB}}{B}\right)^2,
\end{equation}
where $U_{\rm CMB}$ and $U_B$ are the energy densities of the CMB radiation field and the magnetic field, respectively, and 
$B_{\rm CMB} \approx 3.25(1+z)^2\,\mu{\rm G}$ is the equivalent magnetic field strength of the CMB. 
For a representative redshift $z=0.3$, this yields $B_{\rm CMB} \approx 5.5\,\mu{\rm G}$, indicating that inverse-Compton cooling dominates over synchrotron losses throughout much of the cluster outskirts parameter space considered here.

Electrons producing GHz-frequency synchrotron emission are expected to generate IC photons primarily in the hard X-ray band. 
To obtain an order-of-magnitude estimate of the associated flux, we map the synchrotron energy flux $\nu S_{\nu}$ produced by electrons of a given Lorentz factor to the corresponding IC energy flux via
\begin{equation}
\mathcal{E} F_{\rm IC}(\mathcal{E}) \approx
\left(\frac{U_{\rm CMB}}{U_B}\right)\,\nu S_{\nu}.
\end{equation}
To compare with observational limits reported in the soft X-ray band inferred from Suzaku observations of the Toothbrush relic region in the $0.3$--$10~{\rm keV}$ band \citep[e.g.,][]{Itahana2015}, we
estimate the band-integrated IC flux in $0.3$--$10~{\rm keV}$.
In the Thomson regime, IC photons of energy $E_{\rm IC}$ are produced mainly by
electrons with Lorentz factor
\begin{equation}
\mathcal{E}_{\rm IC} \approx \frac{4}{3}\gamma^2 \epsilon_{\rm CMB},
\end{equation}
where $\epsilon_{\rm CMB}$ is a characteristic CMB photon energy.
For $\mathcal{E}_{\rm IC}=0.3$--$10~{\rm keV}$, this corresponds to
$\gamma \sim 6\times10^2$--$3.5\times10^3$, which radiate synchrotron emission at
\begin{equation}
\nu_{\rm syn} \approx \frac{3eB}{4\pi m_ec}\gamma^2
\approx 4.2\,{\rm MHz}\,\left(\frac{B}{1 \mu {\rm G}}\right)\left(\frac{\gamma}{10^3}\right)^2.
\end{equation}
Thus, for $B\sim 1~\mu{\rm G}$, the electrons relevant for $0.3$--$10~{\rm keV}$
IC emission correspond to synchrotron frequencies of order
$\nu_{\rm syn}\sim{\rm a~few}$--$50~{\rm MHz}$, below the commonly observed
$\sim100$--$200~{\rm MHz}$ bands. For an order-of-magnitude estimate we therefore
(i) anchor the normalization to the measured integrated flux density at
$\nu\approx150~{\rm MHz}$ for the Toothbrush relic and (ii) assume that the
low-frequency spectrum is not strongly curved, so that $\nu S_{\nu}$ does not
vary by more than an order of magnitude across the relevant MHz range (e.g.,
for $\alpha\approx1$, $\nu S_{\nu}\propto \nu^{1-\alpha}\approx{\rm const}$).

Adopting $S_{150}$ for the Toothbrush relic, which implies
$\nu S_{\nu}\sim{\rm a~few}\times10^{-15}~{\rm erg\,cm^{-2}\,s^{-1}}$,
and using $z\approx0.3$ (so that $B_{\rm CMB}\approx 3.25(1+z)^2\approx5.5~\mu{\rm G}$),
we obtain a characteristic IC energy flux
$\mathcal{E} F_{\rm IC} \sim (U_{\rm CMB}/U_B)\,\nu S_\nu \sim {\rm a~few} \times 10^{-14}~{\rm erg\,cm^{-2}\,s^{-1}}$
for $B\sim 2~\mu{\rm G}$. Suzaku observations of the Toothbrush relic region report a non-thermal
inverse-Compton upper limit of $2.4\times10^{-13}~{\rm erg\,cm^{-2}\,s^{-1}}$
in the $0.3$--$10~{\rm keV}$ band, implying a lower bound on the magnetic field
strength of $B\gtrsim1.6~\mu{\rm G}$ \citep[][]{Itahana2015}.
This constraint is consistent with our radio-normalized order-of-magnitude estimate.

\section{Summary and Discussion}
In this work, we have examined ballistic surfing acceleration (BSA) as a primary mechanism for producing relativistic electrons in galaxy cluster merger shocks and powering radio relic emission. Our results demonstrate that BSA is a physically viable acceleration channel under cluster conditions characterized by low Mach numbers, weak magnetic fields, and low-level turbulence. Because BSA is governed by large-scale shock electrodynamics rather than stochastic processes, it {\em must operate} regardless of whether other auxiliary acceleration mechanisms are present.

By balancing coherent acceleration by the convection electric field against radiative losses, we find that BSA can accelerate electrons to Lorentz factors $\gamma \sim 10^{4}$--$10^{5}$ under plausible cluster conditions. 
The resulting steady-state electron distributions naturally exhibit high-energy steepening and spectral curvature, producing synchrotron spectra broadly consistent with observed radio relics.
These results indicate that spectral curvature and high-frequency cutoffs in radio relics may arise directly from the interplay between coherent acceleration and radiative cooling, providing a probe of shock electrodynamics in the intracluster medium.

Although uncertainties remain regarding the detailed efficiency of the process, encapsulated in the effective parameter $\eta_{\rm BSA}$, the small values inferred from observations are physically plausible and sufficient to generate the required electron energies owing to the energy-independent acceleration rate of BSA.
Overall, our findings identify BSA as a promising contributor to electron energization in galaxy cluster merger shocks and highlight radio relics as key environments for testing coherent acceleration mechanisms.

\section*{Acknowledgements}
We thank the anonymous referee for helpful comments that improved the quality of the manuscript. This work was supported by the Korea Astronomy and Space Science Institute (KASI). The numerical calculations were performed using computing resources provided by KASI.

\section*{Data Availability}
The observational flux-density data used for comparison in Fig. \ref{fig:f3} are taken from \citet{Stroe2016}. No new observational datasets were generated in this work.

\bibliographystyle{mnras}
\bibliography{bsa_cluster_ref}

\bsp	
\label{lastpage}
\end{document}